\documentclass[aps,prl,twocolumn,showpacs,superscriptaddress]{revtex4}
\usepackage{graphicx}
\usepackage{amssymb}
\usepackage{mathrsfs}
\usepackage{amsmath}
\usepackage{bm,hyperref}
\usepackage{subfigure}
\usepackage{cases}

\begin{document}

\title{Representation of Berry Phase by the Trajectories of Majorana Stars}
\author{H. D. Liu}
\affiliation{National Laboratory of Science and Technology on Computational Physics,
Institute of Applied Physics and Computational Mathematics, Beijing 100088, China}
\affiliation{School of Physics, Northeast Normal University, Changchun 130024, China}
\author{L. B. Fu}
\email[]{lbfu@iapcm.ac.cn}
\affiliation{National Laboratory of Science and Technology on Computational Physics,
Institute of Applied Physics and Computational Mathematics, Beijing 100088, China}\affiliation{HEDPS, Center for Applied Physics and Technology, Peking University, Beijing 100084, China}
\date{\today}

\begin{abstract}
The Majorana's stellar representation, which represents the evolution of a quantum state with
the trajectories of the Majorana stars on a Bloch sphere, provides an intuitive way to study
a physical system with high dimensional projective Hilbert space. In this Letter,
we study the Berry phase by these stars and their loops on the Bloch sphere. It is shown that the Berry phase of a general spin state can be expressed by an elegant formula with the solid angles of Majorana star loops. Furthermore, these results can be naturally used to a general state with arbitrary dimensions. To demonstrate our theory, we study a two mode interacting boson system. Finally, the relation between stars' correlations and quantum entanglement is discussed.
\end{abstract}

\pacs{03.65.Vf, 75.10.Jm, 03.65.Aa, 03.67.Mn}
\maketitle


\paragraph{Introduction.---}
The Berry phase, which reveals the gauge structure associated with cyclic
evolution in Hilbert space\cite{Simon1983}, has become a central
unifying concept of quantum theory\cite{Chruscinski2004,Bohm2003}.
After being introduced into quantum mechanics by Berry\cite{Berry1984}, this phase has been found to play an important role in the study of many important physics
phenomena, such as the quantum Hall effect\cite{Xiao2010,Nagaosa2010}, polarization
of crystal insulators\cite{Kin-Smith1993}, topological phase transition\cite{Xu2011}, and holonomic quantum computation\cite{Loredo2014}. This phase, also known as the
geometric phase, reveals the fact that a quantum eigenstate $|\Psi \rangle $
will acquire an additional geometric phase factor $\oint -\mathrm{Im}\langle
\Psi |\mathrm{d}_{\bm{R}}|\Psi \rangle $ in cyclic adiabatic processes,
where the integral only depends on the geometric path of $\bm{R}$ in the
parameter space.

For the simplest case of an arbitrary two-level state, the geometric path
can be perfectly represented by the close trajectory of a point on the Bloch
sphere, and the Berry phase is proportional to the solid angle subtended by
it. This geometric interpretation seems hard to use for a large spin system
because it is difficult to imagine the trace of a state in the higher
dimensional space. However, Majorana's stellar representation (MSR) builds us a
bridge between the high dimensional projective Hilbert space and the
two-dimensional Bloch sphere \cite{Majorana1932}. In MSR, one can describe a spin-$J$ state (or, equivalently, an $n$
body two-mode boson state with $n=2J$\cite{Schwinger1965}) intuitively by $%
2J $ points on the two-dimensional Bloch sphere rather than one point on a
high dimensional geometric structure, and these $2J$ points are called Majorana
stars (MSs) of the system. This naturally provides an intuitive way to study
the Berry phase for a high spin system\cite{Hannay1998}.

The reason for Majorana's stellar representation drawing much more attention
recently is the studying of spin-orbit coupling in cold atom physics\cite{Kawaguchi2012,Kurn2013}. In
cold atom physics, the large-spin
atoms, such as lanthanide atoms, are introduced as candidates in the process of inducing synthetic gauge field by spin-orbit coupling, since their
narrow linewidth transitions will suppress the additional heating\cite%
{Lian2012,Cui2013}. For high-spin condensates, spin-orbit coupling drives the
Majorana stars moving periodically on the Bloch sphere, i.e., forming the
so-called ``Majorana spin helix." Hence, one will naturally ask, can we have
an explicit relation between the Berry phase and the Majorana stars' helixes or
loops? Recently, Bruno established a novel representation of the Berry phase of
large-spin systems\cite{Bruno2012,Niu2012} by introducing coherent state
representation (CSR) into MSR, and the geometric
phase has been viewed as the Aharonov-Bohm phase acquired by the Majorana
stars as they move through the gas of Dirac strings. However, in Bruno's
excellent work, the connection between the Berry phase and geometric trajectories
of the MSs on the Bloch sphere is still not clear or intuitive. Besides the above research, the MSR has also found
wide applications in various fields. The arrangements and movements of stars have become a powerful tool to study physical problems
related to symmetry, such as classifying the entanglement class\cite{Bastin2009} and computing the
spectrum of the Lipkin-Meshkov-Glick (LMG) model\cite{Ribeiro2007}.

In this Letter, we present a novel formula for the Berry phase of a spin
system which gives an intuitive relation between the Berry phase and MSs'
trajectories on the Bloch sphere. We find that the Berry phase can be decomposed
into two contributions: one is from the sum of the solid angles subtended by
every Majorana star's close trajectory; the other one is from pair correlations
between the stars, which collect the solid angles by the relative motions
of each star pairs. Since any state can be parametrized by the same process of MSR,
these results can naturally be used for any finite quantum system. Using a two-mode boson system, which can be
realized in cold atoms physics(\cite{Cornell2002,Myatt1997}), we calculate
the Berry phase numerically to verify our results. We also find the pair
correlations between stars are naturally related to the quantum
entanglement of the particles. In this respect, it provides an intuitive way
to study measurement and classification of multiparticle entanglement of $n$
particles.

\paragraph{Berry phase in MSR.---}

As we know, a spin-$1/2$ state can be described by a point on the Bloch
sphere. For a spin-$J$ system, its angular momentum operators can be
described by the creation and annihilation operators of two mode bosons with
Schwinger boson representation\cite{Schwinger1965}. Under Schwinger boson representation, the basis of the
spin-$J$ system $|Jm\rangle $ is equivalent to a two mode boson state $%
|J+m,J-m\rangle $.
Therefore a spin-$J$ state $\sum_{-J}^{J}C_{m}|Jm\rangle $ equals to a
generic state of an $n$-dimensional two mode boson system $|\Psi \rangle
^{(n)}=\sum_{-n/2}^{n/2}\frac{C_{m}\hat{a}^{\dag (\frac{n}{2}+m)}\hat{b}%
^{\dag (\frac{n}{2}-m)}}{\sqrt{(\frac{n}{2}+m)!(\frac{n}{2}-m)!}}%
|\varnothing \rangle $ with $n=2J$, which can be factorized as
\begin{eqnarray}
|\Psi \rangle ^{(n)} &=&\frac{1}{N_{n}(\bm{U})}\prod_{k=1}^{n}\hat{a}_{\bm{u}%
_{k}}^{\dag }|\varnothing \rangle  \label{spinjstate} \\
&=&\frac{1}{\sqrt{n!}N_{n}(\bm{U})}\sum_{P}|\bm{u}_{P(1)}\rangle |\bm{u}%
_{P(2)}\rangle \cdots |\bm{u}_{P(n)}\rangle ,  \label{bosonjstate}
\end{eqnarray}%
where $N_{n}(\bm{U})
=[\frac{(n+1)!}{2^{n}}\sum^{[n/2]}_{k=0}\frac{
D^{n}_{k}}{(2k+1)!!}]^{\frac12}$ is the normalization coefficient with $
\bm{U}\equiv \{\bm{u}_{1},\ldots ,\bm{u}_{2J}\}$ (see Supplemental Material\cite{Note1}). The expression of symmetric function
$D^{n}_{k}$\cite{Lee1988} is
$
D^n_k\equiv\sum%
\limits^{n}_{i_1=1}\sum\limits^n_{j_1>i_1}\cdots
\sum\limits^{n*}_{i_k>i_{k-1}}\sum\limits^{n*}_{j_k>i_k}(\bm{u}_{i_1}\cdot%
\bm{u}_{j_1}) \cdots(\bm{u}_{i_k}\cdot\bm{u}_{j_k}),
$
where the $*$
indicates a restriction on the summations so that all nonrepeated indices
in each term take different values. The sum $\sum_{P}$ being over all permutations $P$%
, takes $1,2,\ldots ,n$ to $P(1),P(2),\ldots ,P(n)$. The creation operators
$\hat{a}_{\bm{u}_{k}}^{\dag }\equiv (\cos \frac{\theta _{k}}{2}\hat{a}^{\dag
}+\sin \frac{\theta _{k}}{2}e^{i\phi _{k}}\hat{b}^{\dag })$ and the
annihilation operators $\hat{a}_{\bm{u}_{k}}$ satisfy $[\hat{a}_{\bm{u}%
_{i}}^{\dag },\hat{a}_{\bm{u}_{j}}^{\dag }]=[\hat{a}_{\bm{u}_{i}},\hat{a}_{%
\bm{u}_{j}}]=0$ and $[\hat{a}_{\bm{u}_{i}}^{\dag },\hat{a}_{\bm{u}%
_{j}}]=\langle \bm{u}_{i}|\bm{u}_{j}\rangle $. And $|\bm{u}_{k}\rangle =\cos
\frac{\theta _{k}}{2}\hat{a}^{\dag }|\varnothing \rangle +\sin \frac{\theta
_{k}}{2}e^{i\phi _{k}}\hat{b}^{\dag }|\varnothing \rangle .$ If one denotes $%
\hat{a}^{\dag }|\varnothing \rangle =|\uparrow \rangle $ and $\hat{b}^{\dag
}|\varnothing \rangle =|\downarrow \rangle $ as the orthogonal basis of a
spin-$1/2$ state respectively, then (\ref{bosonjstate}) can be also
understood as a full symmetrized state of $n$ spin-$1/2$ particles\cite%
{Majorana1932}. Consequently, the above factorization will give out $n$
pairs of parameters $\theta _{k},\phi _{k}$ ($k=1,\dots ,n$) which
correspond to $n$ points $\bm{u}_{k}(\theta _{k},\phi _{k})$ on the Bloch
sphere. Therefore, the quantum state in Eq. (\ref{bosonjstate}) and its
evolution can be depicted by these points so-called Majorana stars. Specifically, for the state $|\Psi\rangle^{(n)}$, assuming $x_1,x_2,\dots,x_{n}$ are the roots of
the equation
\begin{equation}
\sum^{n}_{k=0}\frac{(-1)^kC_{n/2-k}}{\sqrt{(n-k)!\,k!}}x^{n-k}=0,
\end{equation}
 then the spherical coordinates $\theta_k$ and $\phi_k$ of $\bm{u}_k$ can
be given by $x_k=\tan\frac{\theta_k}{2}e^{i\phi_k}$\cite{Majorana1932}.

\begin{figure}[b]
\includegraphics*[width=0.95\columnwidth]{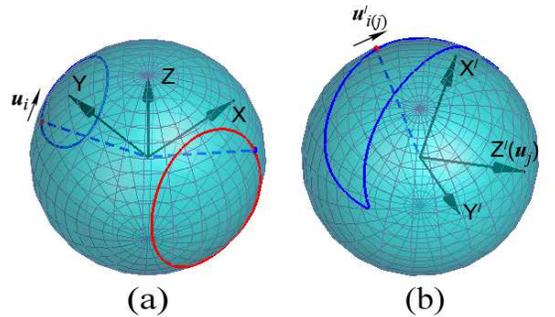}
\caption{(color online) A schematic illustration of (a) the solid angles
subtended by the parallel transports of $\bm{u}_{i}(\theta _{i},\phi _{i})$ and $\bm{u}_{j}(\theta _{j},\phi _{j})$ on the
Bloch sphere: $\Omega _{\bm{u}_{i}}$ (areas subtended by the blue solid
loop) and $\Omega _{\bm{u}_{j}}$ (areas subtended by the red solid
loop); (b) the solid angle in moving frame ($X'(\frac{\protect\pi }{2%
}+\protect\theta _{j},\protect\phi _{j}),Y'(\frac{\protect\pi }{2},-%
\frac{\protect\pi }{2}+\protect\phi _{j}),Z'(\protect\theta _{j},%
\protect\phi _{j})$) subtended by the parallel transports of relative
evolution path between the two stars in (a): $\Omega _{\bm{u}_{i}^{\prime }}$
(areas subtended by the blue solid loop) }
\label{omegad12}
\end{figure}

In particular, for an adiabatic cyclic evolution of the state $|\Psi \rangle
^{(n)}$, each star $\bm{u}_{k}$ traces out an independent loop on the sphere%
\cite{Hannay1998}. As we mentioned, this process will naturally accumulate a
Berry phase for $|\Psi \rangle ^{(n)}$\cite{Berry1984}. Hence, the
interesting task in our scheme, then, is to calculate the Berry phase in
terms of these parametrized loops. According to Berry's definition, the
Berry phase for $|\Psi \rangle ^{(n)}$ reads $\gamma ^{(n)}=\oint
-\mathrm{Im}^{(n)}\langle \Psi |\mathrm{d}_{\bm{u}_{i}}|\Psi \rangle ^{(n)}$
. Substitute Eq. (\ref{bosonjstate}) in it, after a long but straightforward calculation, we find that the contribution of respective evolution of each star can be separated from those of correlations between the stars, and the Berry phase becomes (see the supplemental material\cite{Note1} for details of derivation)
\begin{equation}
\gamma ^{(n)}=\gamma _{0}^{(n)}+\gamma _{C}^{(n)},  \label{gamma}
\end{equation}%
where $\gamma $ can be decomposed into two parts. One part, $\gamma
_{0}^{(n)}=-\sum_{i=1}^{n}\Omega _{\bm{u}_{i}}/2$ is the sum of the solid
angles $\Omega _{\bm{u}_{i}}=\oint (1-\cos \theta _{i})\mathrm{d}\phi _{i}$
subtended by the closed evolution paths of the MSs on the Bloch sphere (as
Fig. \ref{omegad12}-(a) shows).

The another part of Berry phase is
\begin{equation}
\gamma _{C}^{(n)}=\frac{1}{2}\oint \sum_{i=1}^{n}\sum_{j(> i)}^{n}\beta
_{ij}\Omega (\mathrm{d}\bm{u}_{ij}),  \label{gammai}
\end{equation}%
which characterized by the correlations between the stars (hereafter we call
it correlation phase). Here, $\Omega (\mathrm{d}\bm{u}_{ij})\equiv \bm{u}%
_{i}\times \bm{u}_{j}\cdot \mathrm{d}(\bm{u}_{j}-\bm{u}_{i})/d_{ij}$ is the
sum of solid angles of the infinite thin triangle $(\bm{u}_{i},-\bm{u}%
_{j},-\bm{u}_{j}-\mathrm{d}\bm{u}_{j})$ and ($\bm{u}_{j},-\bm{u}_{i},-\bm{u}%
_{i}-\mathrm{d}\bm{u}_{i}$), we denote it as pair solid angle. $\beta _{ij},$
the correlation factor is defined as
\begin{equation}
\beta _{ij}(\bm{D})\equiv -\frac{d_{ij}}{N_{n}^{2}(\bm{D})}\frac{\partial
N_{n}^{2}(\bm{D})}{\partial d_{ij}},  \label{cf}
\end{equation}%
with $\bm{D}=\{d_{ij}\}$ $(i<j)$, in which $d_{ij}\equiv 1-\bm{u}_{i}\cdot %
\bm{u}_{j}$ as the \textquotedblleft distance" between two stars $\bm{u}_{i}(\theta
_{i},\phi _{i})$ and $\bm{u}_{j}(\theta _{j},\phi _{j}).$ Note that the
normalization coefficient $N_{n}^{2}(\bm{U})$ only contains the products of
the first degrees of $d_{ij}$ (see the supplement material\cite{Note1}), and then can be
written as $N_{n}^{2}(\bm{D})=-d_{ij}\frac{\partial N_{n}^{2}(\bm{U})}{%
\partial d_{ij}}+\text{terms without pair}~(\bm{u}_{i},\bm{u}_{j})$.
Therefore, correlation factor $\beta _{ij}(\bm{D})$ is nothing but the
weight of the $d_{ij}$ dependent terms to $N_{n}^{2}(\bm{D})$.
Hence, the correlation phase
can be described as the solid angles between each pair of stars weighted by
their correlation factor $\beta _{ij}$.

Indeed, the pair solid angle $\Omega (\mathrm{d}\bm{u}_{ij})$ can be
expressed by the relative evolutions between $\bm{u}_{i}$ and $\bm{u}_{j}$,
and the absolute evolutions of themselves.  Consider the moving frame in
which the star $\bm{u}_{j}(\theta _{j},\phi _{j})$ is fixed and located at
z-axis $z(0,0)$, the spherical coordinates of the other star $\bm{u}_{i}(\theta
_{i},\phi _{i})$ changes into $\bm{u}_{i(j)}^{\prime }(\theta
_{i(j)}^{\prime },\phi _{i(j)}^{\prime })$ in this frame correspondingly (as Fig.%
\ref{omegad12} shows). On the contrary, we can also obtain the relative
vector $\bm{u}_{j(i)}^{\prime }(\theta _{j(i)}^{\prime },\phi
_{j(i)}^{\prime })$ of $\bm{u}_{j}$ in the moving frame with $\bm{u}%
_{i}(\theta _{i},\phi _{i})$ fixed at z-axis. The pair solid angle $\Omega (%
\mathrm{d}\bm{u}_{ij})$ vector becomes (see Supplement Material\cite{Note1} for details)
\begin{equation}
\Omega(\mathrm{d}\bm{u}_{ij})=[\mathrm{d}%
\phi'_{i(j)}+\mathrm{d}\phi'_{j(i)}]+(\cos\theta_i\mathrm{d}\phi _i+\cos\theta_j\mathrm{d}\phi_j).
 \label{duij}
\end{equation}%
where $\theta ^{\prime }=\theta _{j(i)}^{\prime }=\theta _{i(j)}^{\prime }$
is the angle between $u_{i}$ and $u_{j}$.
Note that the form $\Omega(\mathrm{d}\bm{u})=(1-\cos \theta )\mathrm{d}\phi$ is precisely the
integration element for the solid angle $\Omega _{\bm{u}}$ subtend by the
path of the star $\bm{u}(\theta ,\phi )$. If we integrate Eq. (\ref{duij}), the
geometric meaning of $\Omega (\mathrm{d}\bm{u}_{ij})$ emerges immediately.
Therefore the meaning of
correlation phase $\gamma _{C}^{(n)}$ is quite clear: it consists of the
collection of the weighted relative evolutions between the stars
\begin{equation}
\gamma _{Rij}^{(n)}\equiv \frac{1}{2}\oint \beta _{ij}(\bm{D})\frac{\Omega (%
\mathrm{d}\bm{u}_{i(j)}^{\prime })+\Omega (\mathrm{d}\bm{u}_{j(i)}^{\prime })%
}{1-\bm{u}_{i}\cdot \bm{u}_{j}}
\end{equation}%
with $\Omega (\mathrm{d}\bm{u}_{i(j)}^{\prime })=(1-\cos \theta')\mathrm{d}\phi _{i(j)}^{\prime }$ and the collection of
the weighted absolute evolutions of the pairs of stars
\begin{equation}
\gamma _{Aij}^{(n)}\equiv \frac{1}{2}\oint \beta _{ij}(\bm{D})[\cos \theta
_{i}\mathrm{d}\phi _{i}+\cos \theta _{j}\mathrm{d}\phi _{j}].
\end{equation}%
Namely, $\gamma _{C}^{(n)}=\gamma
_{R}^{(n)}+\gamma _{A}^{(n)}=\sum_{i=1}^{n}\sum_{j(> i)}^{n}(\gamma
_{Rij}^{(n)}+\gamma _{Aij}^{(n)})$. 

So far, we know that the Berry phase in MSR is consist of not only the solid
angles subtended by the paths of the stars but also their correlations.
These results in Eq. (\ref{gamma}) and (\ref{gammai}) are proved to be consist with the marvelous one in Ref. \cite{Bruno2012} which is derived
by introducing the coherent state representation into MSR of the spin-$J$
system.

There follow several notes for some specific cases.

(i) All the stars locate on one single point, i.e. become coincident stars. For this special case $\beta
_{ij}$ and $\Omega (\mathrm{d}u_{ij})$ have value zero, the Berry phase in
Eq. (\ref{gamma}) will be reduced to the sum of solid angles of all stars. This corresponds to the spin coherent state\cite%
{Ganczarek2012,Bruno2012}.

(ii)
This case is just for a spin-$J$ in a uniform magnetic field $\bm{B}=B(\sin
\theta \cos \varphi ,\sin \theta \cos \varphi ,\cos \theta )$. Its
eigenstate $|\Psi \rangle _{m}^{(2J)}=e^{i\hat{J}_{y}\theta }e^{i\hat{J}%
_{z}\varphi }|Jm\rangle $ can be represented by $J+m$ coincident stars $\bm{u}%
(\theta ,\varphi )$ and their $J-m$ coincident antipodal stars $\bm{u}^{\prime
}(\pi -\theta ,\pi +\varphi )$. The Berry phase thus becomes $\gamma
^{(2J)}=\gamma _{0}^{(2J)}=-\frac{1}{2}[(J+m)\Omega _{\bm{u}}-(J-m)\Omega _{%
\bm{u}^{\prime }}]=-m\Omega _{\bm{u}}$, which perfectly matches the
result in Ref.\cite{Berry1990}.

(iii)  All the stars rotate with same angular
velocity as a rigid body. In this case, all the distances between star pairs $\bm{u}_{i}$ and $\bm{u}_{j}$ are
invariant. At this point, $\beta _{ij}$ become constants, and $\gamma
_{C}^{(n)}$ in Eq. (\ref{gammai}) changes into a sum of solid angles as $%
\gamma _{C}^{(n)}=\frac{1}{2}\sum_{i=1}^{n}\sum_{j(\neq i)}^{n}\beta
_{ij}\Omega (\bm{u}_{ij})$, where $\Omega (\bm{u}_{ij})\equiv \oint \Omega (%
\mathrm{d}\bm{u}_{ij})$ is the solid angles accumulated by the infinite
small solid angles $\Omega (\mathrm{d}\bm{u}_{ij})$. By integrating Eq. (\ref%
{duij}), $\Omega (\bm{u}_{ij})$ is turned out to be composed by the solid
angles accumulated by the relative evolution between $\bm{u}_{i}$ and $\bm{u}%
_{j}$ (as Fig.\ref{omegad12}-(b) shows), and the solid angles accumulated by
the evolutions of $\bm{u}_{i}$ and $\bm{u}_{j}$ themselves (as Fig.\ref%
{omegad12}-(a) shows), i.e.
\begin{equation}
\Omega (\bm{u}_{ij})=\frac{\Omega _{\bm{u}_{i(j)}^{\prime }}+\Omega _{\bm{u}%
_{j(i)}^{\prime }}}{1+\bm{u}_{i}\cdot \bm{u}_{j}}-[(\Omega _{\bm{u}%
_{i}}+\Omega _{\bm{u}_{j}})\mod (2\pi )]  \label{cbeta}
\end{equation}%
where $\Omega _{\bm{u}_{i(j)}^{\prime }}$ ($\Omega _{\bm{u}_{j(i)}^{\prime }}
$) is the solid angles subtended by the closed evolution paths of $\bm{u}%
_{i}^{\prime }$ ($\bm{u}_{j}^{\prime }$) relative to $\bm{u}_{j}$ ($\bm{u}%
_{i}$), respectively.

(iv) The pairs of stars $\bm{u}_{i}(\theta _{i},\phi _{i})$ and $\bm{u}_j(\theta _{j},\phi
_{j})$ are always on the same circle of longitude or latitude. The former refers to $\phi
_{i}-\phi _{j}=0,\pm \pi $, i.e. the two stars and the $z$ axis $z(0,0)$ will be
always in the same plane. It will accumulate no loop by the relative
motions between the two stars. For the latter, we have $\theta _{1}=\theta _{2}$ and  the sum of relative
motions between the two stars will also equal to zero owing to the symmetry.
Thus, $\gamma _{Rij}^{(n)}$ will vanish in both of the two situations.
Besides, for $\theta _{i}=\theta _{j}$, if we have $\phi _{i}+\phi _{j}=\text{const}
$, $\gamma _{Aij}^{(n)}$ will also vanish, and this star pair will give
no contribution to the correlation phase.

\paragraph{Two mode interacting boson system.}

To illustrate the above theoretical results, we now consider an interacting boson
system described by Hamiltonian $H=\frac{R\sin \theta }{4}\left( e^{i\varphi
}\hat{a}^{\dag }\hat{b}+e^{-i\varphi }\hat{b}^{\dag }\hat{a}\right) +\frac{%
R\cos \theta }{2}\left( \hat{a}^{\dag }\hat{a}-\hat{b}^{\dag }\hat{b}\right)
+\frac{\lambda }{4}\left( \hat{a}^{\dag }\hat{a}-\hat{b}^{\dag }\hat{b}%
\right) ^{2}$, where $R\cos \theta $ is the energy offset between the two
modes. The parameter $R\sin \theta e^{i\varphi }$ measures the
coupling between the two modes, and $\lambda =g/V$ with $g$ being
the interaction strength between bosons and $V$ being the volume of the system.
This model equals to a spin system\cite{Ribeiro2007} and can be derived from the
bosonic-field Hamiltonian\cite{Cirac1998} and has received extraordinary
attention in the literature on BECs\cite{Leggett2001}. We numerically calculate the parameter dependent eigenstates of the $H$. Then, we use the formula (to avoid errors from numerical differential, see Ref.\cite{Hannay1998}) $e^{i\gamma}=\langle\psi|\psi'\rangle\langle\psi'|\psi''\rangle\cdots\langle\psi^{'\cdots'}|\psi\rangle$ to calculate the Berry phase $\gamma$, and compare it with the numerical result of Eq. (4) in which the time evolutions of majorana stars are obtained by solving the the roots of the equation (3) for corresponding eigenstate numerically.

%

\begin{figure}[tbp]
\includegraphics*[width=\columnwidth]{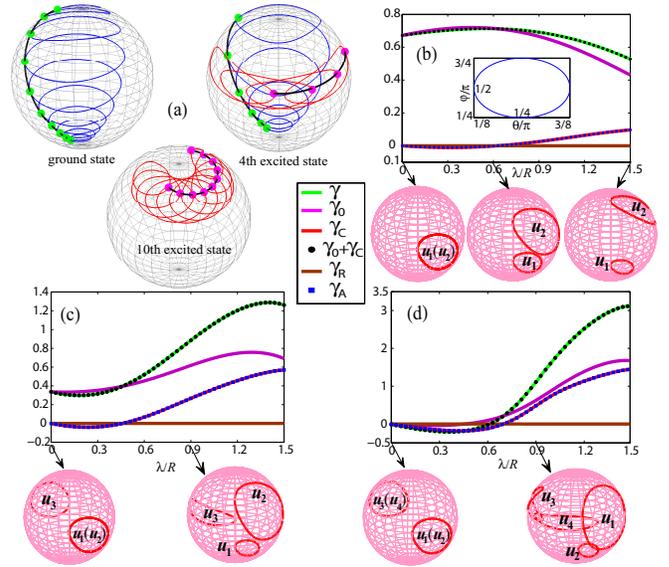}
\caption{(color online) (a) the trajectories and arrangements of stars for the eigenstates of ten bosons with $\lambda/R=0.3$; Dependence of the Berry phase on $\protect%
\lambda/R$ and trajectories of Majorana stars (red loops on spheres) for (b)
ground state of the interaction boson Hamiltonian $H$ with two bosons, (c)
first excited state of three bosons, and (d) second excited state
of four bosons. The insert in (b) shows the evolution of $\theta$ and
$\varphi$ in parameter space.}
\label{example}
\end{figure}

For $\lambda =0$, the
Hamiltonian $H$  reduces to the one of spin-$n/2$ in a magnetic field $\bm{B}=
\frac{R}{2}(\sin\theta\cos\varphi,\sin\theta\cos\varphi,\cos\theta)
$ as in case (ii) above. Therefore, for the $m$th eigenstate, we have $\gamma=(n-2m)\Omega_{\bm{u}}$, for example $2\Omega_{\bm{u}_1}$ for ground state of $H$ with two bosons in Fig. \ref{example} (b),
$\Omega_{\bm{u}_1}$ for first excited state of three bosons in Fig. \ref{example} (c), and $0$ for second excited state of four bosons in Fig. \ref{example} (d), respectively.

As the interacting constant $\lambda $ increase, the
interaction between the bosons breaks the coincidences of stars. Note that
our system can be mapped onto a spin system described by the LMG model similar to Ref.\cite{Ribeiro2007}, the stars for the instantaneous
eigenstate of $H$ are thus spread over two curves on the Bloch sphere. Its $m$th eigenstate has $n+1-m$ stars on one curve
and $m-1$ stars on the other curve\cite{Ribeiro2007}, such as the ground state, $4$th excited state and $10$th excited state for ten bosons shown in Fig. \ref{example}(a).
As the adiabatic parameters evolve, the trajectories of stars become several different
loops and the correlation phases arise. As Fig. \ref{example} shows, the phase $\gamma _{R}^{(n)}$ vanish due to the symmetry between the stars on the two curves. And
the Berry phase $\gamma $, calculated directly by its original
definition, perfectly matches with $\gamma _{0}+\gamma _{C}$ in our
theory.

Besides,
we can use these changes of the
symmetry of the states to clarify the type of state. e.g.
the states of two bosons with coincident stars or separated stars (see the spheres in Fig. \ref{example}(b)); The states
of three bosons with
three coincident stars, two coincident stars or three
separated stars (see the spheres in Fig. \ref{example}(c)). This clarification is according to the
correlation between stars and may be related with entanglement.

\paragraph{Correlations between stars and quantum entanglement.}
In particular, the spin-$J$ state in Eq. (\ref{spinjstate}) is equal to a symmetric $2J$-qubit pure state. By studying the entanglement of
these symmetric qubit pure states, the entanglement of two and
three qubits are found to be determined by the distance $d_{ij}$. For $n=2$, the
concurrence\cite{Wootters1998} equals to $\mathcal{C}=d_{12}/2N_{1}^{2}$. Therefore, the
correlation phase of the state is directly related to its entanglement: $%
\gamma _{C}^{(2)}=\frac{1}{2}\oint \mathcal{C}\Omega (\mathrm{d}\bm{u}_{12})$%
. For $n=3$, there exist different measurements for three different
entanglement classes\cite{Acin2001,Bastin2009} of states: the concurrence for the W type\cite%
{Dur2000} of states becomes $\mathcal{C}_{12}=\frac{2d_{12}}{3N_{3}^{2}}$,
where two of the three stars coincide with each other, and the correlation
phase of the W type state can be written in the form of concurrence like two
qubits: $\gamma _{C}^{(3)}=\frac{3}{2}\oint \Omega (\mathrm{d}\bm{u}_{12})%
\mathcal{C}_{12}$; the $3$-tangle\cite{Coffman2000}
for the Greenberg-Horne-Zeilinger (GHZ)\cite{Greenberg1989} type of states
can be written as $\tau =\frac{2}{3}\beta _{12}\beta _{13}\beta
_{23}N_{3}^{2}$ with three unequal stars; the three same stars at one point bring no entanglement
and thus no correlation phase for the separable states. This means that the types of entanglement can be
distinguished by the number of unequal stars (or diversity degree of the
state\cite{Bastin2009}), and measured by a normalized product of the
distance between unequal stars. Since the classification of entanglement by
the number of unequal stars also hold for $n$ qubits (\cite{Bastin2009}),
such as separable type ($n_{s}=1$), W type ($n_{s}=2$), and GHZ type ($%
n_{s}=n$), the normalized product of distances between unequal stars $%
(\prod_{%
\begin{smallmatrix}
i,j=1 \\
i<j%
\end{smallmatrix}%
}^{n_{s}}d_{ij})/N_{n}^{2(n_{s}-1)}$ may be a valid measure of
entanglement. Besides, there are some other points for MSR worth further studying, such as identical particles in MSR and stars with permuted ends, we will address all these issues in a future paper.

\paragraph{Discussion.}

The Majorana's stellar representation and recently relevant applications have
indicated that the evolution of a high spin state can be displayed intuitively by loops of MSs on the Bloch sphere. Our study here is to show how can
we ``read out" the physical effects of the state such as Berry phase and
entanglement from these stars and loops. The discussion shows that the Berry phase of
a spin-$J$ state is not only determined by the solid angles
subtended by every Majorana star's evolution path but also associated with
the correlation between the stars.
However, if we treat the MSR as a parameterizing process, MSR can be used for any states in a Hilbert space of arbitrary dimensions. For a $n$-dimensional generic state $|\psi\rangle^n=\sum_{m=1}^{n}C_m|m\rangle$, we can still use the roots $y_i=\tan\frac{\theta'_i}{2}e^{i\phi'_i}$  of equation $\sum_{l=0}^{n-1}[(-1)^lC_{n-l}y^{n-1-l}]/(\sqrt{(n-1-l)!l!})=0$
to define $n-1$ Majorana stars $u_i(\theta'_i,\phi'_i)$. These stars can also represent state $|\psi\rangle^n$  accordingly.
Since, the Berry phase is decided by the parameter-dependent probability amplitudes $C_m$ with unchanged basis $|m\rangle$, and the basis $|m\rangle$ of dimension $n$ can be mapped to the basis of state for a spin-$(n-1)/2$ system. The Berry phase for $|\psi\rangle^n$ will take the same form as Eq. (3). Therefore, our results for Berry phase actually hold for any finite quantum system, and can be widely applied in various fields.

\ \ \\
We thank B. Wu and H. Zhai for helpful discussions. This work is supported by the National Basic Research Program of China (973 Program) (Grants No. 2013CBA01502, No. 2011CB921503, and No. 2013CB834100), the National Natural Science Foundation of China (Grants No. 11374040, No. 11274051, and No. 11405008).

\end{document}


\title{Supplement Material: Representation of Berry phase for spin system by the trajectories of Majorana stars}
\author{H. D. Liu}
\affiliation{National Laboratory of Science and Technology on Computational Physics,
Institute of Applied Physics and Computational Mathematics, Beijing 100088, China}

\author{L. B. Fu\footnote{fu\_libin@iapcm.ac.cn}}
\affiliation{National Laboratory of Science and Technology on Computational Physics,
Institute of Applied Physics and Computational Mathematics, Beijing 100088, China}\affiliation{HEDPS, Center for Applied Physics and Technology, Peking University, Beijing 100084, China}
\maketitle

\section{I. Derivation of the Normalization Coefficient $N_n^2$}
To prove the expression of $N_n^2$, we first notice that the states of MS $|\bm{u}_k\rangle$ have exchange symmetry in Eq. (2) in the letter. Therefore, its normalized coefficient can be calculated by picking any single star to interact with other stars
\begin{equation}
\begin{aligned}
N_n^2=&\left(\langle\bm{u}_1|\langle\bm{u}_2|\cdots\langle\bm{u}_{n}|\right)\sum_{P}|\bm{u}_{P(1)}\rangle|\bm{u}_{P(2)}\rangle\cdots|\bm{u}_{P(n)}\rangle\\
=&\langle\bm{u}_i|\bm{u}_i\rangle (N_n)_i'^2\\
&+\sum_{j(\neq i)}^{n}\langle\bm{u}_i|\bm{u}_j\rangle\langle\bm{u}_j|\bm{u}_i\rangle (N_n)'^2_{ij}\\
&+\sum_{j,k(\neq i,j)}\langle\bm{u}_i|\bm{u}_j\rangle\langle\bm{u}_j|\bm{u}_k\rangle\langle\bm{u}_k|\bm{u}_i\rangle (N_n)'^2_{ijk}\\
&+\cdots\\
&+\langle\bm{u}_i|\left(\sum_{P}|\bm{u}_{P(1)}\rangle\langle\bm{u}_{P(1)}|\cdots|\bm{u}_{P(n-2)}\rangle\langle\bm{u}_{P(n-2)}|\right)'_{i}|\bm{u}_i\rangle,
\end{aligned}
\label{qubitN}
\end{equation}
where The notation $'_{i}$, $'_{ij}$, $'_{ijk}$ indicates that removing star $\bm{u}_i$, ($\bm{u}_i$ and $\bm{u}_j$) and ($\bm{u}_i$, $\bm{u}_j$ and $\bm{u}_k$) from the product in the first line of Eq. (\ref{qubitN}), respectively. Notice that, $|\bm{u}_i\rangle\langle\bm{u}_i|=(1+\sigma\cdot\bm{u}_i)/2$ and the relation $(\bm{\sigma}\cdot\bm{u}_i)(\bm{\sigma}\cdot\bm{u}_j)=\bm{u}_i\cdot
\bm{u}_j+i\bm{\sigma}\cdot(\bm{u}_i\times\bm{u}_j)$. For exchange symmetry, the item with cross product will vanish in the expression of normalization coefficient and leave the items with the product of $\bm{u}_i\cdot\bm{u}_j$.
Next, we use mathematical induction to prove the expression
\begin{equation}
N_n(\bm{U})=(n!\sum\limits_P\prod\limits_k\langle\bm{u}_k|\bm{u}%
_{P(k)}\rangle )^{\frac12}=[\frac{(n+1)!}{2^{n}}\sum^{[n/2]}_{k=0}\frac{%
D^{n}_{k}}{(2k+1)!!}]^{\frac12}
\label{nc}
\end{equation}
where, he expression of symmetric function
$D^{n}_{k}$ is
$
D^n_k\equiv\sum%
\limits^{n}_{i_1=1}\sum\limits^n_{j_1>i_1}\cdots
\sum\limits^{n*}_{i_k>i_{k-1}}\sum\limits^{n*}_{j_k>i_k}(\bm{u}_{i_1}\cdot%
\bm{u}_{j_1}) \cdots(\bm{u}_{i_k}\cdot\bm{u}_{j_k}),
$
and the $*$
indicates a restriction on the summations so that all non-repeated indices
in each term take different values. The sum $\sum_{P}$ being over all permutations $P$%
, taking $1,2,\ldots ,n$ to $P(1),P(2),\ldots ,P(n)$.

For $n=1$, the state reduce to $|\bm{u}\rangle$ and $N_n(\bm{U})=\langle\bm{u}|\bm{u}\rangle=1$. The expression (\ref{nc}) holds.

Assuming Eq. (\ref{nc}) holds for natural number $n$. For the situation of $n+1$, we can add a new star $\bm{u}_{n+1}$ to the state
$|\Psi \rangle^{(n)}$. Then Eq. (\ref{qubitN}) becomes
\begin{eqnarray}
N_{n+1}^2=&&\left(\langle\bm{u}_1|\langle\bm{u}_2|\cdots\langle\bm{u}_{n}\langle\bm{u}_{n+1}|\right)\sum_{P}|\bm{u}_{P(1)}\rangle|\bm{u}_{P(2)}\rangle\cdots|\bm{u}_{P(n+1)}\rangle\nonumber\\
=&&\langle\bm{u}_{n+1}|\bm{u}_{n+1}\rangle N_n^2\nonumber\\
&&+\sum_{i=1}^{n}\langle\bm{u}_{n+1}|\bm{u}_i\rangle\langle\bm{u}_i|\bm{u}_{n+1}\rangle (N_n)'^2_{i}\nonumber\\
&&+\sum_{i,j(\neq i)}\langle\bm{u}_{n+1}|\bm{u}_i\rangle\langle\bm{u}_i|\bm{u}_j\rangle\langle\bm{u}_j|\bm{u}_{n+1}\rangle (N_n)'^2_{ij}\nonumber\\
&&+\cdots\nonumber\\
&&+\langle\bm{u}_{n+1}|\bm{u}_i\rangle\langle\bm{u}_i|\left(\sum_{P}|\bm{u}_{P(1)}\rangle\langle\bm{u}_{P(1)}|\cdots|\bm{u}_{P(n-2)}\rangle\langle\bm{u}_{P(n-2)}|\right)'_{i}|\bm{u}_{n+1}\rangle\nonumber\\
=&&N_n^2\nonumber\\
&&+\sum_{i}^{n}\langle\bm{u}_i|\left(\frac{1+\bm{\sigma}\cdot\bm{u}_{n+1}}{2}\right)|\bm{u}_i\rangle (N_n)'^2_{i}\nonumber\\
&&+\sum_{i,j(\neq i)}\langle\bm{u}_j|\left(\frac{1+\bm{\sigma}\cdot\bm{u}_{n+1}}{2}\right)|\bm{u}_i\rangle\langle\bm{u}_i|\bm{u}_j\rangle (N_n)'^2_{ij}\nonumber\\
&&+\cdots\nonumber\\
&&+\langle\bm{u}_i|\left(\sum_{P}|\bm{u}_{P(1)}\rangle\langle\bm{u}_{P(1)}|\cdots|\bm{u}_{P(n-2)}\rangle\langle\bm{u}_{P(n-2)}|\right)'_{i}\left(\frac{1+\bm{\sigma}\cdot\bm{u}_{n+1}}{2}\right)|\bm{u}_i\rangle\nonumber\\
=&&N_n^2+\frac12\sum_{i=1}^{n}\left[\langle\bm{u}_i|\bm{u}_i\rangle (N_n)'^2_{i}+\sum_{j(\neq i)}\langle\bm{u}_j|\bm{u}_i\rangle\langle\bm{u}_i|\bm{u}_j\rangle (N_n)'^2_{ij}+\cdots+\langle\bm{u}_i|\left(\sum_{P}\prod_{l=1}^{n}|\bm{u}_{P(l)}\rangle\langle\bm{u}_{P(l)}|\right)'_{i}|\bm{u}_i\rangle\right]\nonumber\\
&&+\sum_{i}^{n}\left(\frac{\bm{u}_{n+1}\cdot\bm{u}_i}{2}\right)| (N_n)'^2_{i}\nonumber\\
&&+\sum_{i,j(\neq i)}|\left(\frac{\bm{u}_{n+1}\cdot\bm{u}_i+\bm{u}_{n+1}\cdot\bm{u}_j}{4}\right)(N_n)'^2_{ij}\nonumber\\
&&+\cdots\nonumber\\
&&+\langle\bm{u}_i|\left(\sum_{P}|\bm{u}_{P(1)}\rangle\langle\bm{u}_{P(1)}|\cdots|\bm{u}_{P(n-2)}\rangle\langle\bm{u}_{P(n-2)}|\right)'_{i}\left(\frac{\bm{\sigma}\cdot\bm{u}_{n+1}}{2}\right)|\bm{u}_i\rangle.
\label{ntn1}
\end{eqnarray}
After a straightforward calculation, the contribution of $\bm{u}_{n+1}$ to the products of $l$ pair like $\bm{u}_{i}\cdot\bm{u}_j$ is
\begin{equation}
\frac{(n-2l+1)!l!}{(2l-1)!2^{n-l+1}}\sum_{m=2l-1}^n\frac{m!(n-m+1)}{(m-2l+1)!}=\frac{(n+2)!l!}{2^{n-l+1}(2l+1)!}=\frac{(n+2)!}{2^{n+1}(2l+1)!!}.
\label{lc}
\end{equation}
Substituting Eq. (\ref{qubitN}) and (\ref{lc}) into Eq. (\ref{ntn1}), we finally have
\begin{equation}
\begin{aligned}
N_{n+1}=&\left(1+\frac n2\right)N_n\\
&+\frac{(n+2)!}{3\times2^{n+1}}\sum_{i}^{n}\left(\bm{u}_{n+1}\cdot\bm{u}_i\right)\\
&+\frac{(n+2)!}{15\times2^{n+1}}\sum_{i,j,k}(\bm{u}_{n+1}\cdot\bm{u}_i)(\bm{u}_j\cdot\bm{u}_k)(N_n)'^2_{ij}~~~~~~(\text{the sum restrict to non-repeated indices as in $D_k^n$})\\
&+\cdots\\
=&\frac{(n+2)!}{2^{n+1}}\sum^{[(n+1)/2]}_{k=0}\frac{%
D^{n+1}_{k}}{(2k+1)!!}.
\end{aligned}
\end{equation}
Thus, the expression hold for $n+1$ and Eq. (\ref{qubitN}) has been proved.
\section{II. Derivation of the Berry phase $\gamma^{(n)}$}
In particular, for an adiabatic cyclic evolution of state $|\Psi \rangle
^{(n)}$, each star $\bm{u}_{k}$ traces out an independent loop on the sphere. As we mentioned, this process
will naturally accumulate a Berry phase for $|\Psi \rangle ^{(n)}$. According to Berry's definition, the Berry phase for $|\Psi
\rangle ^{(n)}$ reads: $\gamma ^{(n)}=\oint A(\bm{U})
=-\mathrm{Im}^{(n)}\langle \Psi |\mathrm{d}_{\bm{u}_{i}}|\Psi \rangle ^{(n)}$,
and $A(\bm{U})$ takes the form
\begin{equation}
\begin{aligned}
{^n\langle}\Psi|\mathrm{d}_{\bm{u}_{i}}\Psi\rangle^n=&\sum_{i=1}^{n}\frac{(N^2_n)_i'\langle\bm{u}_i|d\bm{u}_i\rangle}{N^2_n}\\
&+\sum_{
\begin{smallmatrix}
i,j=1\\
i\neq j
\end{smallmatrix}}^{n}\frac{
\langle\bm{u}_i|d\bm{u}_j\rangle\langle\bm{u}_j|\bm{u}_i\rangle (N^2_n)'_{ij}}{N^2_n}\\
&+\sum_{
\begin{smallmatrix}
i,j=1\\
i\neq j
\end{smallmatrix}}^{n}\sum_{k(\neq i,j)}\frac{
\langle\bm{u}_i|d\bm{u}_j\rangle\langle\bm{u}_j|\bm{u}_k\rangle\langle\bm{u}_k|\bm{u}_i\rangle (N^2_n)'_{ijk}}{N^2_n}\\
&+\cdots\\
&+\sum_{
\begin{smallmatrix}
i,j=1\\
i\neq j
\end{smallmatrix}}^{n}\frac{
\langle\bm{u}_i|d\bm{u}_j\rangle\langle\bm{u}_j|}{N^2_n}\left(\sum_{P}\prod_{l=1}^{n}|\bm{u}_{P(l)}\rangle\langle\bm{u}_{P(l)}|\right)'_{ij}|\bm{u}_i\rangle.
\end{aligned}
\label{dqubit}
\end{equation}
Compare with Eq. (\ref{qubitN}), the differential element $\langle\bm{u}_i|\mathrm{d}\bm{u}_j\rangle$ are important for the calculation of Eq. (\ref{dqubit}). Consider a infinity small cyclic product of stars. By calculating at the leading order, we have
\begin{equation}
\begin{aligned}
\langle\bm{u}_i|\bm{u}_j+d\bm{u}_j\rangle\langle\bm{u}_j+d\bm{u}_j|\bm{u}_j\rangle\langle\bm{u}_j|\bm{u}_i\rangle
&=\langle\bm{u}_i|d\bm{u}_j\rangle\langle\bm{u}_j|\bm{u}_i\rangle-
\langle\bm{u}_i|\bm{u}_j\rangle\langle\bm{u}_j|d\bm{u}_j\rangle\langle\bm{u}_j|\bm{u}_i\rangle+
\langle\bm{u}_i|\bm{u}_j\rangle\langle\bm{u}_j|\bm{u}_i\rangle\\
&=\frac{2+2\bm{u}_i\cdot\bm{u}_j+(\bm{u}_i+\bm{u}_j)\cdot d\bm{u}_j-i\bm{u}_i\times\bm{u}_j\cdot\mathrm{d}\bm{u}_j}{4},
\end{aligned}
\end{equation}
notice that $|\langle\bm{u}_i|\bm{u}_j\rangle|^2=\frac{1+\bm{u}_i\cdot\bm{u}_j}{2}$,
then we obtain
\begin{equation}
\langle\bm{u}_i|d\bm{u}_j\rangle=\langle\bm{u}_i|\bm{u}_j\rangle\left[\langle\bm{u}_j|d\bm{u}_j\rangle+\frac{(\bm{u}_i+\bm{u}_j)\cdot \mathrm{d}\bm{u}_j}{2(1+\bm{u}_i\cdot\bm{u}_j)}-\frac{i\bm{u}_i\times\bm{u}_j\cdot \mathrm{d}\bm{u}_j}{2(1+\bm{u}_i\cdot\bm{u}_j)}\right].
\label{dij}
\end{equation}
Substituting (\ref{dij}) into Eq. (\ref{dqubit}), and using the relations $(\bm{\sigma}\cdot\bm{A})(\bm{\sigma}\cdot\bm{B})=\bm{A}\cdot
\bm{B}+i\bm{\sigma}\cdot(\bm{A}\times\bm{B})$ and $(\bm{A}\times\bm{B})\cdot(\bm{C}\times\bm{D})=(\bm{A}\cdot\bm{C})(\bm{B}\cdot\bm{D})-(\bm{A}\cdot\bm{D})(\bm{B}\cdot\bm{C})$.
The imaginary part of Eq. (\ref{dqubit}) can be calculated directly as
\begin{eqnarray}
\mathrm{Im}{^n\langle}\Psi|\mathrm{d}_{\bm{u}_i}\Psi\rangle^n=&&\sum_{i=1}^{n}\frac{(N^2_n)_i'\mathrm{Im}\langle\bm{u}_i|d\bm{u}_i\rangle}{N^2_n}\nonumber\\
&&+\mathrm{Im}\sum_{
\begin{smallmatrix}
i,j=1\\
i\neq j
\end{smallmatrix}}^{n}
\left[\langle\bm{u}_j|d\bm{u}_j\rangle+\frac{(\bm{u}_i+\bm{u}_j)\cdot \mathrm{d}\bm{u}_j}{2(1+\bm{u}_i\cdot\bm{u}_j)}-\frac{i\bm{u}_i\times\bm{u}_j\cdot \mathrm{d}\bm{u}_j}{2(1+\bm{u}_i\cdot\bm{u}_j)}\right]\cdot\nonumber\\
&&\frac{
\langle\bm{u}_i|\bm{u}_j\rangle}{N^2_n}\left[\langle\bm{u}_j|\bm{u}_i\rangle (N^2_n)'_{ij}
+\sum_{k(\neq i,j)}\langle\bm{u}_j|\bm{u}_k\rangle\langle\bm{u}_k|\bm{u}_i\rangle (N^2_n)'_{ijk}
+\cdots
+\langle\bm{u}_j|\left(\sum_{P}\prod_{l=1}^{n}|\bm{u}_{P(l)}\rangle\langle\bm{u}_{P(l)}|\right)'_{ij}|\bm{u}_i\rangle\right]\nonumber\\
=&&\sum_{i=1}^{n}\frac{\mathrm{Im}\langle\bm{u}_i|d\bm{u}_i\rangle}{N^2_n}\left[(N^2_n)_i'
+\sum_{j\neq i}\langle\bm{u}_i|\bm{u}_j\rangle\langle\bm{u}_j|\bm{u}_i\rangle (N^2_n)'_{ij}
+\cdots
+\langle\bm{u}_i|\bm{u}_j\rangle\langle\bm{u}_j|\left(\sum_{P}\prod_{l=1}^{n}|\bm{u}_{P(l)}\rangle\langle\bm{u}_{P(l)}|\right)'_{ij}|\bm{u}_i\rangle\right]\nonumber\\
&&+\mathrm{Im}\sum_{
\begin{smallmatrix}
i,j=1\\
i\neq j
\end{smallmatrix}}^{n}
\left[\frac{(\bm{u}_i+\bm{u}_j)\cdot \mathrm{d}\bm{u}_j}{2(1+\bm{u}_i\cdot\bm{u}_j)}-\frac{i\bm{u}_i\times\bm{u}_j\cdot \mathrm{d}\bm{u}_j}{2(1+\bm{u}_i\cdot\bm{u}_j)}\right]\nonumber\\
&&\cdot\mathrm{Tr}\left\{\frac{(1+\bm{\sigma}\cdot\bm{u}_i)(1+\bm{\sigma}\cdot\bm{u}_j)}{4N^2_n}
\left[(N^2_n)'_{ij}+\sum_{k(\neq i,j)}
\frac{(1+\bm{\sigma}\cdot\bm{u}_k)(N^2_n)'_{ijk}}{2}
+\cdots+\left(\sum_{P}\prod_{l=1}^{n}\frac{1
+\bm{\sigma}\cdot\bm{u}_{P(l)}}{2}\right)'_{ij}
\right]
\right\}\nonumber\\
=&&\sum_{i=1}^{n}\mathrm{Im}\langle\bm{u}_i|\mathrm{d}\bm{u}_i\rangle\nonumber\\
&&+\mathrm{Im}\sum_{
\begin{smallmatrix}
i,j=1\nonumber\\
i\neq j
\end{smallmatrix}}^{n}\left[\frac{(\bm{u}_i+\bm{u}_j)\cdot \mathrm{d}\bm{u}_j}{2(1+\bm{u}_i\cdot\bm{u}_j)}-\frac{i\bm{u}_i\times\bm{u}_j\cdot \mathrm{d}\bm{u}_j}{2(1+\bm{u}_i\cdot\bm{u}_j)}\right]\mathrm{Tr}\left\{\frac{1+\bm{\sigma}(\bm{u}_i+\bm{u}_j)+\bm{u}_i\cdot\bm{u}_j+i\bm{\sigma}\cdot(\bm{u}_i\times\bm{u}_j)}{4N^2_n}\cdot\right.\nonumber\\
&&\left.
\left[(N^2_n)'_{ij}+\sum_{k(\neq i,j)}
\frac{(1+\bm{\sigma}\cdot\bm{u}_k)(N^2_n)'_{ijk}}{2}
+\cdots+\left(\sum_{P}\prod_{l=1}^{n}\frac{1
+\bm{\sigma}\cdot\bm{u}_{P(l)}}{2}\right)'_{ij}
\right]
\right\}\nonumber\\
=&&\sum_{i=1}^{n}\mathrm{Im}\langle\bm{u}_i|\mathrm{d}\bm{u}_i\rangle\nonumber\\
&&+\frac{1}{N^2_n}\sum_{
\begin{smallmatrix}
i,j=1\\
i\neq j
\end{smallmatrix}}^{n}\mathrm{Tr}\left\{\left[\frac{\bm\sigma\cdot(\bm{u}_i\times\bm{u}_j)(\bm{u}_i+\bm{u}_j)\cdot \mathrm{d}\bm{u}_j}{8(1+\bm{u}_i\cdot\bm{u}_j)}-\frac{(1+\bm{u}_i\cdot\bm{u}_j+\bm{\sigma}\cdot(\bm{u}_i+\bm{u}_j))\bm{u}_i\times\bm{u}_j\cdot \mathrm{d}\bm{u}_j}{8(1+\bm{u}_i\cdot\bm{u}_j)}\right]\right.\nonumber\\
&&\left.\cdot\left[(N^2_n)'_{ij}+\sum_{k(\neq i,j)}
\frac{(1+\bm{\sigma}\cdot\bm{u}_k)(N^2_n)'_{ijk}}{2}
+\cdots+\left(\sum_{P}\frac{\prod_{l=1}^{n}(1
+\bm{\sigma}\cdot\bm{u}_{P(l)})}{2}\right)'_{ij}
\right]
\right\}\nonumber\\
=&&\sum_{i=1}^{n}\mathrm{Im}\langle\bm{u}_i|\mathrm{d}\bm{u}_i\rangle\nonumber\\
&&-\frac{1}{N^2_n}\sum_{
\begin{smallmatrix}
i,j=1\\
i\neq j
\end{smallmatrix}}^{n}\frac{\bm{u}_i\times\bm{u}_j\cdot \mathrm{d}\bm{u}_j}{8}\mathrm{Tr}\left[(N^2)'_{ij}+\sum_{k(\neq i,j)}
\frac{(1+\bm{\sigma}\cdot\bm{u}_k)(N^2)'_{ijk}}{2}
+\cdots+\left(\sum_{P}\frac{\prod_{l=1}^{n}(1
+\bm{\sigma}\cdot\bm{u}_{P(l)})}{2}\right)'_{ij}
\right]\nonumber\\
=&&\sum_{i=1}^{n}\mathrm{Im}\langle\bm{u}_i|\mathrm{d}\bm{u}_i\rangle+\frac12\sum_{
\begin{smallmatrix}
i,j=1\nonumber\\
i\neq j
\end{smallmatrix}}^{n}\frac{\bm{u}_i\times\bm{u}_j\cdot \mathrm{d}\bm{u}_j}{N^2_n}\frac{\partial N^2_n}{\partial d_{ij}}\nonumber\\
=&&\sum_{i=1}^{n}\mathrm{Im}\langle\bm{u}_i|\mathrm{d}\bm{u}_i\rangle+\frac12\sum_{
\begin{smallmatrix}
i,j=1\nonumber\\
i< j
\end{smallmatrix}}^{n}\frac{\bm{u}_i\times\bm{u}_j\cdot \mathrm{d}(\bm{u}_j-\bm{u}_i)}{N^2_n}\frac{\partial N^2_n}{\partial d_{ij}}
\label{ddqubit}
\end{eqnarray}
where

\begin{equation}
\langle\bm{u}_j|\mathrm{d}\bm{u}_j\rangle=i\left(\frac{1-\cos\theta_j}{2}\right)d\phi_j,
\label{djj}
\end{equation}
with spherical 	
coordinates $ \theta_j$ and $\phi_j$ of $\bm{u}_j$. And
\begin{equation}
\begin{aligned}
\frac{\partial N^2_n}{\partial d_{ij}}&\equiv-\mathrm{Tr}\left\{\frac14\left[N'^2_{ij}+\sum_{k(\neq i,j)}
\frac{(1+\bm{\sigma}\cdot\bm{u}_k)N'^2_{ijk}}{2}
+\cdots+\left(\sum_{P}\frac{\prod_{l=1}^{n}(1
+\bm{\sigma}\cdot\bm{u}_{P(l)})}{2}\right)'_{ij}
\right]\right\}\\
&=-\frac{(n+1)!}{2^{n}}\sum^{[n/2]-1}_{k=0}\frac{1}{(2k+3)!!}(D^{n-2}_k)'_{ij},
\label{nd}
\end{aligned}
\end{equation}
with the distance $d_{ij}=1-\bm{u}_1\cdot\bm{u}_2$. Here we notice that although Eq. (1) in the letter is a symmetric state where the MSs possess the exchange symmetry, and the spin-$J$ system in MSR can be treated as a $2J$-boson system, it still differs from the regular identical boson system. It is because the states $|\bm{u}_i\rangle$ are nonorthogonal and thus $|\Psi\rangle^J$ is unnormalized. This is the reason why we have the term $\frac{\partial N^2_n}{\partial d_{ij}}$ which is not exist in the regular identical boson system.
Compare Eq. (\ref{nc}) with Eq. (\ref{nd}), we find that
\begin{equation}
N^2_n=d_{ij}\frac{\partial N^2_n}{\partial d_{ij}}+ \text{terms without pair} (\bm{u}_i,\bm{u}_j)
\end{equation}
By substituting Eq. (\ref{ddqubit}) into the definition of $A(\bm{u})$,  the Berry phase becomes
\begin{equation}
\gamma ^{(n)}=\gamma _{0}^{(n)}+\gamma _{C}^{(n)},  \label{gamma}
\end{equation}%
where

\begin{equation}
\gamma
_{0}^{(n)}=-\sum_{i=1}^{n}\Omega _{\bm{u}_{i}}/2
\end{equation}
 is the collection of the
solid angles $\Omega _{\bm{u}_{i}}=\oint (1-\cos \theta _{i})\mathrm{d}\phi
_{i}$ of the closed evolution paths of the MSs on the Bloch sphere. And
\begin{equation}
\gamma _{C}^{(n)}=\frac{1}{2}\oint \sum_{i=1}^{n}\sum_{j(>i)}^{n}\beta
_{ij}(\bm{D})\Omega (\mathrm{d}\bm{u}_{ij}),  \label{gammai}
\end{equation}
with the correlation factor
\begin{equation}
\beta _{ij}(\bm{D})\equiv -\frac{d_{ij}}{N_{n}^{2}(\bm{D})}\frac{\partial
N_{n}^{2}(\bm{D})}{\partial d_{ij}}.  \label{cf}
\end{equation}%
and the pair solid angle
\begin{equation}
\Omega (\mathrm{d}\bm{u}_{ij})\equiv \bm{u}_{i}\times %
\bm{u}_{j}\cdot \mathrm{d}(\bm{u}_{j}-\bm{u}_{i})/d_{ij}
\end{equation}
which is the sum of
solid angles of the infinitely thin triangle $(\bm{u}_{i},-\bm{u}_{j},-\bm{u}%
_{j}-\mathrm{d}\bm{u}_{j})$ and ($\bm{u}_{j},-\bm{u}_{i},-\bm{u}_{i}-\mathrm{%
d}\bm{u}_{i}$).

\section{III. Derivation of solid angle $\Omega(\mathrm{d}\mathbf{u}_{ij})$ in Eq. (7) in the letter }
Suppose we rotate $\bm{u}_i(\theta_i,\phi_i)=(\sin\theta\cos\phi,\sin\theta\sin\phi,\cos\theta)$ to $\bm{z}(0,0)$  by
\begin{equation}
T_i=\left( \begin{matrix}
   \cos\theta_i & 0 & -\sin\theta_i  \\
   0 & 1 & 0  \\
   \sin\theta_i & 0 & \cos\theta_i  \\
\end{matrix} \right)\left( \begin{matrix}
   \cos\phi_i & \sin\phi_i & 0  \\
   -\sin\phi_i & \cos\phi_i & 0  \\
   0 & 0 & 1  \\
\end{matrix} \right),
\end{equation}
$\bm{u}_j(\theta_j,\phi_j)$ will change into
\begin{equation}
\bm{u}%
_{j(i)}^{\prime }(\theta _{j(i)}^{\prime },\phi _{j(i)}^{\prime })=\left(-\cos\theta_2\sin\theta_1+\sin\theta_2\cos\theta_1\cos(\phi_1-\phi_2),-\sin\theta_2\sin(\phi_1-\phi_2),\cos\theta'\right). \end{equation}
where $\theta'\equiv\theta_{j(i)}^{\prime }=\arccos[\cos\theta_i\cos\theta_j+\sin\theta_i\sin\theta_j\cos(\phi_i-\phi_j)]$ is the angle
between $\bm{u}_i$ and $\bm{u}_j$, and
\begin{equation}
\phi _{j(i)}^{\prime}=\arctan\left[\frac{-\sin\theta_j\sin(\phi_i-\phi_j)}{-\cos\theta_j\sin\theta_i+\sin\theta_j\cos\theta_i\cos(\phi_i-\phi_j)}\right]
.\label{uji}
\end{equation}
The vector calculation in $\Omega(\mathrm{d}\bm{u}_{ij})$ becomes
\begin{equation}
\begin{aligned}
\bm{u}_i\times\bm{u}_j\cdot\mathrm{d}\bm{u}_j&=[(T_i^{-1}\bm{z})\times(T_i^{-1}\bm{u}
_{j(i)}^{\prime })]\cdot\mathrm{d}(T_i^{-1}\bm{u}
_{j(i)}^{\prime })\\
&=T_i^{-1}(\bm{z}\times\bm{u}
_{j(i)}^{\prime })\cdot (T_i^{-1}\mathrm{d}\bm{u}
_{j(i)}^{\prime })+T_i^{-1}(\bm{z}\times\bm{u}
_{j(i)}^{\prime })\cdot (\mathrm{d}T_i^{-1}\bm{u}
_{j(i)}^{\prime })\\
&=\bm{z}\times\bm{u}
_{j(i)}^{\prime }\cdot \mathrm{d}\bm{u}'_j+\bm{z}\times\bm{u}
_{j(i)}^{\prime }\cdot (T_1\mathrm{d}T_1^{-1}\bm{u}
_{j(i)}^{\prime })
\end{aligned}
\label{uiz}
\end{equation}
Similarly, the rotation
\begin{equation}
T_j=\left( \begin{matrix}
   \cos\theta_j & 0 & -\sin\theta_j  \\
   0 & 1 & 0  \\
   \sin\theta_j & 0 & \cos\theta_j  \\
\end{matrix} \right)\left( \begin{matrix}
   \cos\phi_j & \sin\phi_j & 0  \\
   -\sin\phi_j & \cos\phi_j & 0  \\
   0 & 0 & 1  \\
\end{matrix} \right),
\end{equation}
will change $\bm{u}_j$ and $\bm{u}_i$ into $\bm{z}(0,0)$ and
\begin{equation}
\bm{u}
_{i(j)}^{\prime }(\theta _{i(j)}^{\prime },\phi _{i(j)}^{\prime })=\left(-\cos\theta_i\sin\theta_j+\sin\theta_i\cos\theta_j\cos(\phi_j-\phi_i),-\sin\theta_i\sin(\phi_j-\phi_i),\cos\theta'\right). \label{uij}
\end{equation}
respectively, where $\theta _{i(j)}^{\prime }=\theta'$, and
\begin{equation}
\phi _{i(j)}^{\prime }=\arctan\left[\frac{-\sin\theta_i\sin(\phi_j-\phi_i)}{-\cos\theta_i\sin\theta_j+\sin\theta_i\cos\theta_j\cos(\phi_j-\phi_i)}\right]
\end{equation}
Then, we have
\begin{equation}
\bm{u}_j\times\bm{u}_i\cdot\mathrm{d}\bm{u}_i=\bm{z}\times\bm{u}
_{i(j)}^{\prime }\cdot \mathrm{d}\bm{u}
_{i(j)}^{\prime }+\bm{z}\times\bm{u}
_{i(j)}^{\prime }\cdot (T_j\mathrm{d}T_j^{-1}\bm{u}
_{i(j)}^{\prime }).
\label{ujz}
\end{equation}
Substituting Eq. (\ref{uji}), (\ref{uiz}), (\ref{uij}) and (\ref{ujz}) into the definition of $\Omega(\mathrm{d}\bm{u}_{ij})$, and notice the distance $d_{ij}=1-\bm{u}_i\cdot\bm{u}_j$ is invariant under the two rotations, we have
\begin{equation}
\begin{aligned}
\Omega(\mathrm{d}\bm{u}_{ij})&=\frac{\bm{u}_i\times\bm{u}_j\cdot\mathrm{d}\bm{u}_j+\bm{u}_j\times\bm{u}_i\cdot\mathrm{d}\bm{u}_i}{1-\bm{u}_i\cdot\bm{u}_j}\\
&=\frac{\bm{z}\times\bm{u}
_{j(i)}^{\prime }\cdot \mathrm{d}\bm{u}'_{j(i)}+\bm{z}\times\bm{u}
_{j(i)}^{\prime }\cdot (T_1\mathrm{d}T_1^{-1}\bm{u}
_{j(i)}^{\prime })+\bm{z}\times\bm{u}
_{i(j)}^{\prime }\cdot \mathrm{d}\bm{u}
_{i(j)}^{\prime }+\bm{z}\times\bm{u}
_{i(j)}^{\prime }\cdot (T_j\mathrm{d}T_j^{-1}\bm{u}
_{i(j)}^{\prime })}{1-\bm{u}_i\cdot\bm{u}_j}\\
&=[\mathrm{d}
\phi'_{i(j)}+\mathrm{d}\phi'_{j(i)}]+(\cos\theta_i\mathrm{d}\phi _i+\cos\theta_j\mathrm{d}\phi_j)\\
&=\frac{(\cos\theta_1-\cos\theta_2)(d\phi_2-d\phi_1)+(\sin\theta_1d\theta_2-\sin\theta_2d\theta_1)\sin(\phi_1-\phi_2)}{1-\cos\theta'}
+(\cos\theta_i\mathrm{d}\phi _i+\cos\theta_j\mathrm{d}\phi_j).
\end{aligned}
\label{duij}
\end{equation}